\documentstyle[12pt]{article}
\newcommand{\be}{\begin{equation}}
\newcommand{\ee}{\end{equation}}
\newcommand{\bea}{\begin{eqnarray}}
\newcommand{\eea}{\end{eqnarray}}

\newcommand{\eps}{\varepsilon}
\newcommand{\SUT}{SU(2)}
\newcommand{\SUN}{SU(N)}
\newcommand{\ZN}{Z_N}
\newcommand{\NN}{{\cal N}}
\newcommand{\Amu}{A_{\mu}(x)}

\newcommand{\half}{{\scriptstyle{{1\over 2}}}}

\newcommand{\quart}{{\scriptstyle{{1\over 4}}}}

\newcommand{\sixteenth}{{\scriptstyle{{1\over 16}}}}

\newcommand{\real}{\relax{\rm I\kern-.18em R}}
\newcommand{\Pexp}{{\rm P\,exp}}
\newcommand{\Tr}{{\rm Tr}}

\newcommand{\TTxR}{T^3\times R}

\newcommand{\La}{\Lambda}

\def\ba{\begin{array}}
\def\ea{\end{array}}
\begin{document}
\vskip-1cm
\hfill FTUAM-99-13; IFT-UAM/CSIC-99-19
\vskip0mm
\hfill INLO-PUB-9/99
\vskip5mm
\begin{center}
{\LARGE{\bf{\underline{Nahm dualities on the torus - a synthesis}}}}\\
\vspace*{7mm}{\large
Margarita Garc\'{\i}a P\'erez$^{(a)}$, Antonio Gonz\'alez-Arroyo$^{(a,b)}$,\\
Carlos Pena$^{(a)}$ and Pierre van Baal$^{(c)}$\\}
\vspace*{5mm}
($a$) Departamento de F\'{\i}sica Te\'{o}rica C-XI,\\
Universidad Aut\'{o}noma de Madrid,\\
28049 Madrid, Spain.\\
\vspace*{3mm}
($b$) Instituto de F\'{\i}sica Te\'{o}rica C-XVI,\\
Universidad Aut\'{o}noma de Madrid,\\
28049 Madrid, Spain.\\
\vspace*{3mm}
($c$) Instituut-Lorentz for Theoretical Physics,\\
University of Leiden, PO Box 9506,\\
NL-2300 RA Leiden, The Netherlands.
\end{center}
\vspace*{2mm}{\narrower{\noindent
\underline{Abstract:}
We give a unified description of self-dual $\SUT$ gauge fields on tori of 
size $\l_t \times l_s^3$, based on a mixture of analytical and numerical 
methods using the Nahm transformation, extended to the case of twisted 
boundary conditions. We show how torus calorons ($l_t/l_s$ small) are 
Nahm dual to the torus instantons ($l_t/l_s$ large). Holonomies are 
dual to the locations of constituents, this duality becoming exact in the 
limiting cases $l_s$ or $l_t\rightarrow\infty$. Implications for the moduli 
spaces are discussed.
}\par}

\section{Introduction}
The study of $\SUN$ gauge fields on the torus has a long history (see
Ref.~\cite{RevT} for a recent account). It was initiated by G. `t
Hooft~\cite{Tho1} who pointed out additional non-trivial topological features, 
associated to what is known as {\em twisted boundary conditions} (tbc), 
describing the presence of center-valued conserved fluxes for non-abelian 
gauge fields. In particular, the study of self-dual gauge field configurations 
on the torus has become a challenging problem. Both within the finite volume 
Hamiltonian and the finite temperature formulations of non-abelian gauge 
theories they play an important role, and the main result of this paper is 
a rather surprising detailed dual relationship between these solutions.

For non-zero twist, numerical methods based on the lattice formulation of 
gauge theories strongly suggested the existence of non-abelian self-dual 
solutions for all torus sizes~\cite{MTB,MaTo}. Indeed, existence of smooth 
solutions with one unit of topological charge ($Q$) and any non-trivial twist 
has been established rigorously~\cite{BrMT}, even though one can prove 
non-existence for zero twist~\cite{BrvB}. (For higher topological charges 
existence has been proven by Taubes~\cite{Tau} earlier). To date, however, 
no analytic expressions have been found for these non-abelian solutions on 
the 4-torus. This contrasts with the situation for $R^4$ (or $S^4$), where 
one has the algebraic ADHM construction~\cite{ADHM}. 

An important technique for studying self-dual gauge fields is the Nahm 
transformation~\cite{Nahm}, which is particularly simple for the torus. It maps 
$\SUN$ self-dual solutions of charge $Q$ on the torus to $SU(Q)$ self-dual 
solutions of charge $N$ on the dual torus. The mapping is an involution,
i.e. applying the Nahm transformation again brings one back to the original 
gauge field configuration. The only analytically known example of the Nahm 
transformation on the torus, based on a particular abelian 
solution~\cite{Tho2}, maps it onto itself~\cite{Buck}. Nevertheless, the Nahm 
transformation is believed to be an important ingredient in leading to 
analytical results for the self-dual solutions. 

A few strategies have been used to improve our understanding of this problem.
One follows from the consideration of mixed compact and non-compact directions 
$T^n\times R^{4-n}$. Under the duality transformation the non-compact 
directions collapse and the dual space is an $n$-dimensional torus, thereby 
achieving a dimensional reduction. The most dramatic example is that of $n=0$, 
where the dual space collapses to a point, being the reason that the ADHM 
construction~\cite{ADHM} is algebraic. The $n=1$ case is relevant at finite 
temperature~\cite{HaSh,GPY} for which the associated geometry is 
$R^3\times S^1$. The case $n=3$ was considered as a way to circumvent the 
no-go theorem on $Q=1$, zero-twist configurations~\cite{MaSn,MaPi}. This 
geometry is relevant for the Hamiltonian formulation of gauge fields on the 
spatial torus, with the non-compact direction identified with time. The 
possibility of having $Q=1$ configurations introduces an important additional 
simplification, since its Nahm-dual connection is an abelian gauge field.

Applying Nahm's transformation in the non-compact case demands some
modifications~\cite{Buck}. In particular, there appear a finite number of 
singularities where the Nahm transformed gauge field is non-selfdual. These 
occur when the holonomy of the original self-dual gauge field (extended to 
$U(N)$ by adding a curvature free abelian gauge field, which parametrises 
the coordinates of the dual torus) has a trivial eigenvalue. The holonomy is 
defined by the Wilson loops winding non-trivially in the compact directions, 
when taken to infinity in the non-compact directions. 

An important example of the appearance of these singularities in the Nahm 
transformation is the case of calorons~\cite{HaSh,GPY}, instantons at finite 
temperature. It was already known from the early work of Nahm~\cite{Nahm} 
that the dual formulation involves a $U(1)$ gauge field on the circle with 
suitable singularities. It was only recently that explicit expressions for 
the $Q=1$ self-dual configurations on $R^3\times S^1$ with {\em non-trivial} 
holonomy were obtained~\cite{ThPi,Lee}. In particular it was shown how the 
location of the singularities in the dual field is related to the holonomy 
(the Polyakov loop at spatial infinity). These non-trivial calorons reveal 
the presence of {\em constituent BPS monopoles}, hidden at trivial holonomy 
due the masslessness of one of the monopoles. 

Another example recently analysed in detail is the $Q=1$ instanton 
configuration for $\TTxR$ with twisted boundary conditions in time~\cite{Pie3}.
There the location of the singularities is determined by the holonomies 
(Polyakov loops in the three directions of $T^3$) at infinite time. Away from 
these singularities, the {\em abelian} Nahm transformed field is self-dual and 
therefore satisfies Maxwell equations. 
The singularities act as sources and can be interpreted as point-like dyons 
carrying (equal) electric and magnetic charge.

It is interesting to ask oneself how these caloron and instanton
configurations are approached from configurations on a torus, $l_t\times l_s^3
$, when sending either $l_s$ or $l_t$ to infinity. When finite, these configurations
will be referred to as torus calorons and torus instantons respectively.
Both types of configurations have been analysed in the past by means 
of numerical methods. In~\cite{MaSn} it was seen how some of the $\TTxR$
configurations are the limit of self-dual configurations on the 4-torus with
twisted boundary conditions in time, which are known to exist~\cite{BrMT}.
The $Q=1$ solution 
tunnels between vacua where some of the holonomies have opposite signs.
The non-existence of the $Q=1$ solution without twist was 
understood as an obstruction for tunnelling between two vacua with the same 
holonomy. The situation seems quite similar to the O(3) model on the cylinder, 
which turned out to be tractable analytically~\cite{Snip}. 

The case of non-zero spatial twist was studied numerically even earlier. The 
basic building block is a lump --the twisted instanton-- carrying fractional 
topological charge ($Q=\half$ for $\SUT$) and which is well localised in 
time~\cite{MTB,MaTo,MTAC}. Single twisted instanton configurations arise for 
non-zero (so called non-orthogonal) twist, both in space and time, and they 
have a well-defined limit as $l_t\rightarrow\infty$. Only their overall 
position is a free parameter, so that in particular their scale is fixed 
(to $l_s$). Higher topological charge configurations look as superpositions of 
these twisted instantons, distributed in the time-direction~\cite{RTNcol,TPA}. 
These $Q=\half$ lumps also feature prominently in numerical studies that 
address how 4-torus configurations approach the non-compact geometries of 
$T^2\times R^2$~\cite{T2R2}.

Finally, caloron configurations on a torus have been recently 
studied~\cite{Calo} on lattices typically having $l_s/l_t=4$ or higher. 
These torus calorons fitted extremely well to the infinite volume analytic 
results. Interestingly, twist in the time direction enforces the holonomy 
(when taking the limit $l_s\rightarrow\infty$) to have zero trace, which 
implies the constituent monopoles of the caloron have equal mass. For spatial 
twists, however, the numerical analysis of Ref.~\cite{Calo} suggests that 
the constituent monopoles are forced to sit at fixed relative positions on 
the torus, but can have arbitrary masses (only constrained by the total 
topological charge). One can also study one constituent monopole in isolation 
by having suitable twists, both in space and time, such that it supports the 
$Q=\half$ twisted instanton mentioned before. Indeed in the limit 
$l_t/l_s\rightarrow 0$ this solution becomes time independent and thus has 
to approach a single BPS monopole. Hence also in this limit the twisted 
instanton - as the BPS monopole - is the basic building block, but now 
arbitrarily placed in space.

Quite remarkably, as will be discussed in this paper, torus calorons are 
dual to torus instantons, with either their space (for calorons) or time 
(for instantons) positions determined by the holonomies of their Nahm dual. 
In the limit $l_t\rightarrow\infty$, relevant for $\TTxR$, the Nahm dual 
has the limit $\hat l_t\rightarrow 0$ (setting $l_s=1=1/l_s$ for convenience). 
This means that the non-abelian cores of the torus caloron shrink to zero 
and one is left with the abelian field discussed before. Quite in the same 
way one can interpret the abelian dual of the caloron in the limit
$l_s\rightarrow\infty$ (here setting units by $l_t=1$, the Nahm dual 
has $\hat l_s\rightarrow 0$) as the limit in
which the instanton cores have shrunk to zero. In both cases the non-abelian
field is self-dual at all values of $l_t$ and $l_s$, and the violation of the
abelian self-duality in the limiting case is thus due to ignoring the singular
non-abelian cores of the constituents. Such understanding of the singularity
may prove important for future analytic work in the $\TTxR$ case.

In our analysis, two ingredients play an important part. One is that for 
$T^4$ lattice techniques have been developed~\cite{Nlat} that allow us to 
study the Nahm transformation numerically. This was tested in a non-trivial 
example for $\SUT$ with charge 2, which maps to itself. This demonstrated 
that the method is precise enough to become a useful tool in numerically 
analysing the properties of the Nahm dualities. This charge 2 configuration 
was actually composed from four $Q=\half$ twisted instantons. As we will see,
it is these twisted instantons that are mapped on to themselves as they only 
have their position as a free parameter.

The other ingredient is the recent formulation of the Nahm transformation in 
the presence of twisted boundary conditions~\cite{Ntbc}. The Nahm transform 
of an $SU(N)$, charge $Q$ self-dual configuration is an $SU(N_0Q)$ 
configuration with topological charge $\hat{Q}=N/N_0$, where $N_0$ is an 
integer depending on the twist. If the original twist is non-trivial the 
Nahm transform lives on a torus with non trivial twist. In essence the 
construction is related to recognising twist as (for $\SUT$) half-periods 
for a larger torus, to which the usual Nahm transformation can be applied. 
Remarkably, the dual configuration admits ``half-periods'' as well, and the 
dual gauge field can thus be formulated in terms of gauge fields with twisted 
boundary conditions. It is this that allows us to demonstrate that the 
$Q=\half$ $\SUT$ twisted instanton is mapped to itself under the Nahm 
transformation. 

Alternative to the formulation in terms of ``doubling'', one can introduce
a suitable flavour group to compensate for the twist~\cite{Flav}, with which 
the Nahm transformation can then be performed. Again one might study 
half-periods in the resulting dual torus and reach a self-dual configuration 
in a smaller torus (the Nahm-dual torus) and with non-zero twist (the Nahm-dual
twist)~\cite{Ntbc}. Furthermore, it was shown that both formulations agree. 
{}From the analytical point of view this method is more general and allows a 
simple and compact characterisation of the Nahm-dual torus and twist. This 
will be presented in an Appendix. 

In this paper we will make a synthesis of all the results mentioned above,
leading to additional insight in self-dual gauge fields on the torus and their 
moduli spaces. In Section 2 we will set up the formalism, whereas section 3 
shows how the torus calorons and torus instantons are related under the Nahm 
transformation, illustrated explicitly by numerical examples. We show that 
taking the appropriate limits, as discussed qualitatively in this introduction,
reproduces the known analytic results. 

\section{The Nahm transformation with twist}
Let us consider a 4 dimensional torus. This can be defined as
$R^4/\Lambda$ where $\Lambda$ is a rank-four lattice of points. Let
$e^{(\alpha)}$ stand for four vectors ($\alpha \in \{0,1,2,3\}$) which can be
chosen as generators of $\Lambda$.  We will take in general
a hypercubic torus $l_t\times l_1\times l_2\times l_3$, for which 
$e^{(0)}=(l_t,0,0,0)$, $e^{(1)}=(0,l_1,0,0)$, $e^{(2)}=(0,0,l_2,0)$ and 
$e^{(3)}=(0,0,0,l_3)$. Often we will take $l_1=l_2=l_3\equiv l_s$. 
For obvious physical reasons we will refer to the direction with length 
$l_t$ as time and to the other three as space. 

Now consider $\SUN$ self-dual gauge fields $\Amu$ defined on this torus. 
They can be looked at as fields defined on $R^4$, periodic modulo gauge
transformations under translations by vectors $a$ in the lattice $\Lambda$.
Given any element of $\Lambda$,  $a\equiv e(s)=s_{\alpha}e^{(\alpha)}$ 
($s_\alpha\in Z$), we have (our convention throughout this paper is that 
$\Amu$ is hermitian)
\be
A_{\nu}(x+a)=[\Omega_a]A_{\nu}(x)\equiv\Omega_a(x)\,A_{\nu}(x)\,
\Omega^{\dagger}_a(x)+i\ \Omega_a(x)\,\partial_{\nu}\Omega^{\dagger}_a(x)\quad,
\ee
where $\Omega_a(x)$ form a family of $SU(N)$ matrices, known as {\em twist 
matrices}. Since the lattice $\Lambda$ is abelian, the matrices $\Omega_a(x)$ 
must satisfy the following consistency conditions
\be
\label{tbc1}
\Omega_a(x+b)\,\Omega_b(x)=e^{2\pi i\NN(a,b)}\,\Omega_b(x+a)\,\Omega_a(x)
\quad. 
\ee
Because the matrices belong to $\SUN$, the factor $\exp(2\pi i\,\NN(a,b))$
must be an element of the center $\ZN$. Hence, $\NN(a,b)$ is an 
antisymmetric bilinear form defined by its action on the generators
\be
\NN(e^{(\alpha)},e^{(\beta)})=\frac{n_{\alpha \beta }}{N}\quad,
\ee
where $n_{\alpha \beta}$ is an antisymmetric matrix of integers defined
modulo $N$, known as the twist tensor. We will frequently refer to its
elements in the form of two 3-vectors of integers: $k_i=n_{0i}$ and
$m_i=\half\eps_{ijk}n_{jk}$. 

An important role in what follows will be played by the lattice $\Lambda_0$, a 
sublattice of $\La$. This is given by the elements  $a\in \La$, such that the
associated gauge transformation $\Omega_a(x)$ commutes (in the sense of the 
composition in Eq.~(\ref{tbc1})) with all other $\Omega_b(x)$. Equivalently
\be
\label{la_zero}
 \Lambda_0   \equiv  \{ a \in \La \; |\;  \NN(a,b)\in Z,\quad \forall b\in \La
\}\quad . 
\ee
The quotient group $\La/\Lambda_0$ is a finite abelian group. Using the 
freedom to select a basis of $\La$ such that $n_{\alpha \beta }$ has the 
canonical Frobenius form, one can show that the order of this group is the 
square of some integer, $N_0^2$. The integer $N_0$ plays an 
important role in what follows and depends  only on the form $\NN$. It is 
clear that for trivial twist vectors $\vec{k}=\vec{m}=\vec{0} \bmod N$ 
($\NN(a,b) \in Z$) one has $\Lambda_0=\La$ and $N_0=1$.

Now let us proceed to introduce Nahm's transformation~\cite{Nahm,BrvB}. 
For that purpose add a 4 parameter family of curvature free abelian
gauge fields to the original self-dual gauge field, $\Amu+2\pi z_{\mu}I$.
This family of $U(N)$ gauge fields is still self-dual, as the curvature
is not affected by the addition. The zero-modes of the Weyl equation in 
this background satisfy  
\be
\label{weyl}
(\overline{D}-2\pi i\bar{z})\Psi_z^{\alpha}(x)=0\quad.
\ee
In this notation we associate to each four-vector $C_\mu$ a $2\times2$ matrix
$\overline{C}=C_0\,I_{2\times 2}+i\vec{C}\cdot\vec{\sigma}$ (with $\sigma_k$ 
the Pauli matrices). The covariant derivative $D_{\mu}$ is given by 
$\partial_{\mu}-i\Amu$. In the absence of twist ($\vec{k}=\vec{m}=0$), the 
Weyl spinors satisfy the boundary conditions:
\be 
\label{boundary}
\Psi_z^{\alpha}(x+a)= \Omega_a(x) \Psi_z^{\alpha}(x)\quad,
\ee
and the index theorem~\cite{Index} implies that the number of solutions
satisfying Eqs.~(\ref{weyl}-\ref{boundary}) (labelled  by $\alpha$) is given by the topological
charge $Q$. Then the Nahm transformed gauge field is given by
\be
\label{NA}
\hat A_\mu^{\alpha\beta}(z)= i\int d^4 x\,\Psi_z^\alpha(x)^\dagger
\frac{\partial}{\partial z_\mu}\Psi_z^{\beta}(x)\quad. 
\ee
Quite miraculously this is an $SU(Q)$ self-dual gauge field with topological
charge $N$. Furthermore, the gauge field is defined on a dual torus
$\tilde R^4/\tilde\Lambda$,  where  $\tilde\Lambda$ is the dual lattice of 
$\Lambda$.  

With non-trivial twist ($\vec{k},\vec{m}\ne 0$) the boundary conditions shown 
in Eq.~(\ref{boundary}) are no longer valid, since $\Psi_z^\alpha(x)$ takes 
values in the fundamental representation of $\SUN$, which transforms 
non-trivially under the center of the gauge group. Two strategies were
developed in Ref.~\cite{Ntbc} to deal with this. One involves adding
flavour and the other replicating the original torus. Both methods were 
shown to lead to the same Nahm transformed gauge field.

For the {\em flavour} construction one introduces the $U(N_fN)$ self-dual
gauge field $A_{\mu}(x)\otimes\,I_{N_f\times N_f}$. The enlargement of the 
colour space allows us to modify the twist matrices as follows: 
$\Omega_a(x)\rightarrow\Omega_a(x)\otimes\Gamma^{*}(a)$. The
$N_f\times N_f$ constant matrices $\Gamma(a)$ satisfy
\be
\label{twisteaters}
\Gamma(a)\,\Gamma(a')=e^{2\pi i\NN(a,a')}\,\Gamma(a')\,\Gamma(a)\quad.
\ee
In this way the new $U(N_fN)$ matrices $\Omega_a(x)$ have no twist, and the
standard formalism can be applied to them. The existence of solutions to
Eq.~(\ref{twisteaters}), known as twist-eaters, was studied in the
past~\cite{TwEat}. The constant matrices $\Gamma(a)$ are known to exist
provided $N_f$ is a multiple of $N_0$, the twist-dependent integer defined
below Eq.~(\ref{la_zero}). The solutions form an irreducible set provided the
matrices are precisely $N_0\times N_0$. Furthermore, the solution is unique
modulo similarity transformations and multiplication by
constants~\cite{TwEat,RevT}. If we restrict to matrices belonging to
$U(N_0)$ these constants are necessarily phases (a $U(1)$ representation
of the lattice $\La$). Hence, we will consider $N_f=N_0$ and make a particular
choice of solutions $\Gamma(a)$. In the appendix we  show how different  choices
of the twist eaters correspond to translations in $z$ of the Nahm transformed  field.

We have reduced the problem  to the situation  without twist, for a $U(N_0 N)$ 
self-dual gauge field  defined on the torus $R^4/\Lambda$ with topological 
charge $\hat{N}=N_0Q$. One can then construct the Nahm transformed gauge field 
in the standard way. It will be a $U(\hat N)$ self-dual gauge field with 
topological charge $N_0N$, living on the dual torus $R^4/\tilde\Lambda$.
The construction is in terms of the solutions of the Weyl equation in the 
background of the $U(N_0 N)$ gauge field (adding $2\pi z_\mu I$). However, 
given the tensor product form of this gauge field, these  zero-modes are seen 
to be solutions of the original $U(N)$ Weyl equation~(\ref{weyl}). But the 
boundary condition, Eq.~(\ref{boundary}), is replaced by
\be
\label{boundary2}
\Psi_z^{i\alpha}(x+a)=\Omega_a(x)\,\Gamma_{ij}^*(a)\,\Psi_z^{j\alpha}(x)\quad ,
\ee
where $i,j$ are flavour indices (taking $N_0$ values). The index theorem tells 
us that the label $\alpha$, running over the linearly independent solutions, 
takes $\hat{N}=N_0Q$ values. Choosing   these solutions to form an orthonormal 
set, we can express the dual gauge field as follows:
\be
\label{NT}
\hat A_\mu^{\alpha \beta}(z)=i\sum_{i=1}^{N_0}\int d^4 x\,\Psi_z^{i\alpha}
(x)^\dagger\frac{\partial}{\partial z_{\mu}}\Psi_z^{i\beta}(x)\quad.
\ee
Although this field is in principle living on the torus $R^4/\tilde{\La}$, in 
Ref.~\cite{Ntbc} it is shown  that it possesses additional periodicities. 
Hence, one can define the Nahm transform  of the original gauge field with 
non-trivial twist as the restriction of the gauge field given in Eq.~(\ref{NT})
to the minimal torus, in which case it has non-trivial twist as well. It can 
be shown  that this minimal torus is precisely  $R^4/\hat{\La}$, where 
$\hat{\La}=\tilde\La_0$, the dual lattice of $\Lambda_0$ (see Eq. (\ref{la_zero})). This definition 
preserves the property that the Nahm transformation is an involution, and reduces 
to the standard construction for trivial twist. Proofs are given in 
Ref.~\cite{Ntbc}. A more elegant, basis independent derivation is given 
in the appendix. 

Now we will explain briefly what is the essence of the alternative 
construction, which is  directly related to our numerical implementation of
the Nahm transform. We can view $\Amu$ on $R^4/\Lambda$ as being defined on 
the 4 dimensional hypercube formed by the unit cell (spanned by $e^{(\mu)}$) 
of $\Lambda$, with appropriated twisted boundary conditions. Consider now a 
sublattice $\Lambda'$ whose unit cell is obtained by duplicating the original 
hypercube in various directions, in such a way that the net twist vanishes. 
So one defines $\Amu$ on $R^4/\Lambda'$, such that its boundary conditions 
do no longer carry twist. A trivial way to achieve this is by taking 
$\Lambda'=\Lambda_0$, whose unit cell contains $N_0^2$ unit cells of 
$\Lambda$, but this is not a minimal choice.

To determine $\Lambda'$ it is simplest to make a choice of generators  
for $\Lambda$ such that  $n_{\alpha\beta}$ takes the  Frobenius standard form with
non-zero entries defined by $n_{02}=-n_{20}=q_1$ and $n_{31}=-n_{13}=q_2$.
We introduce $p_1$ 
and $p_2$ as the {\em smallest} positive integers such that $q_ip_i/N\in Z$. 
One duplicates the hypercube $(p_1-1)$ times in {\em either} the 0 or 2 
direction and $(p_2-1)$ times in {\em either} the 1 or 3 direction such 
that $N_0=p_1p_2$. Further freedom arises in case $p_i$ has non-trivial 
prime factors. The new gauge field on $R^4/\Lambda'$ has topological charge 
$N_0Q$ and no twist. Now one can apply the Nahm transformation, which maps 
to a $SU(N_0Q)$ self-dual gauge field of topological charge $N$, defined 
on $R^4/\tilde\Lambda'$. 

\begin{figure}[htb]
\vskip5.2cm
\includegraphics{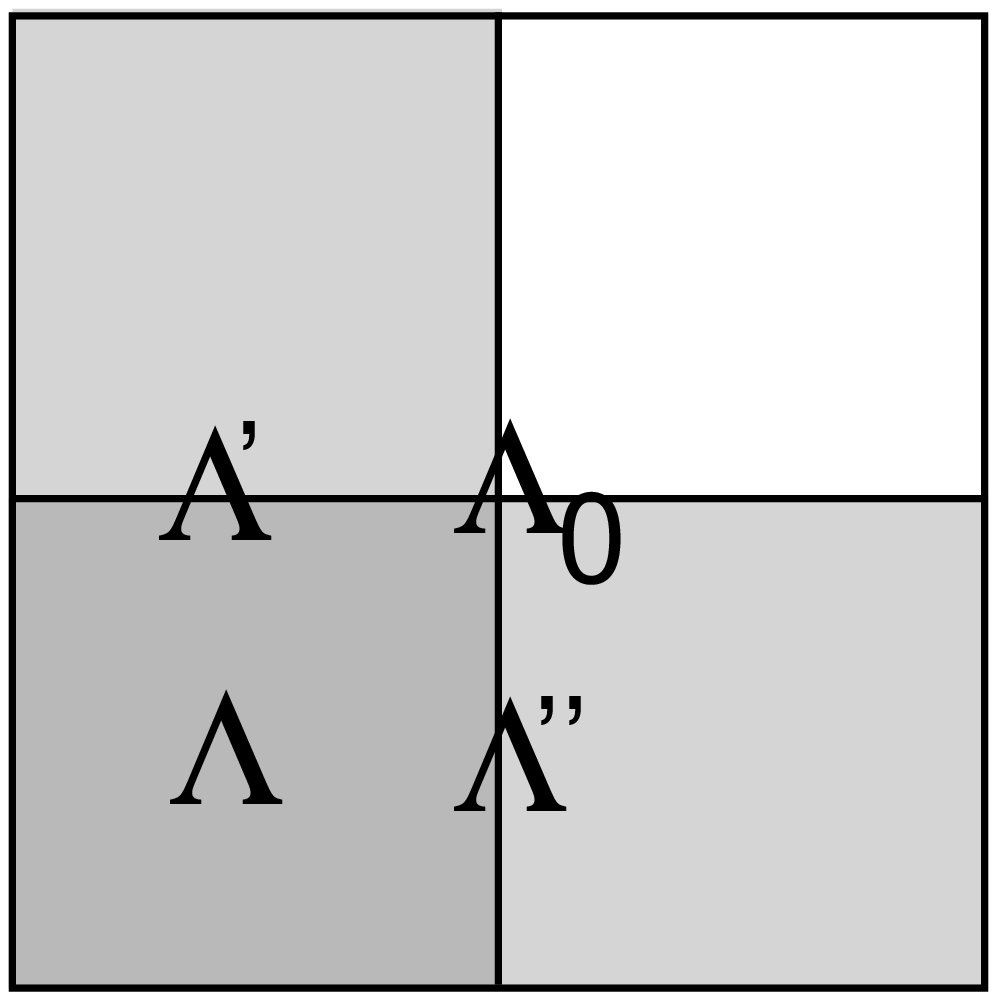}
\includegraphics{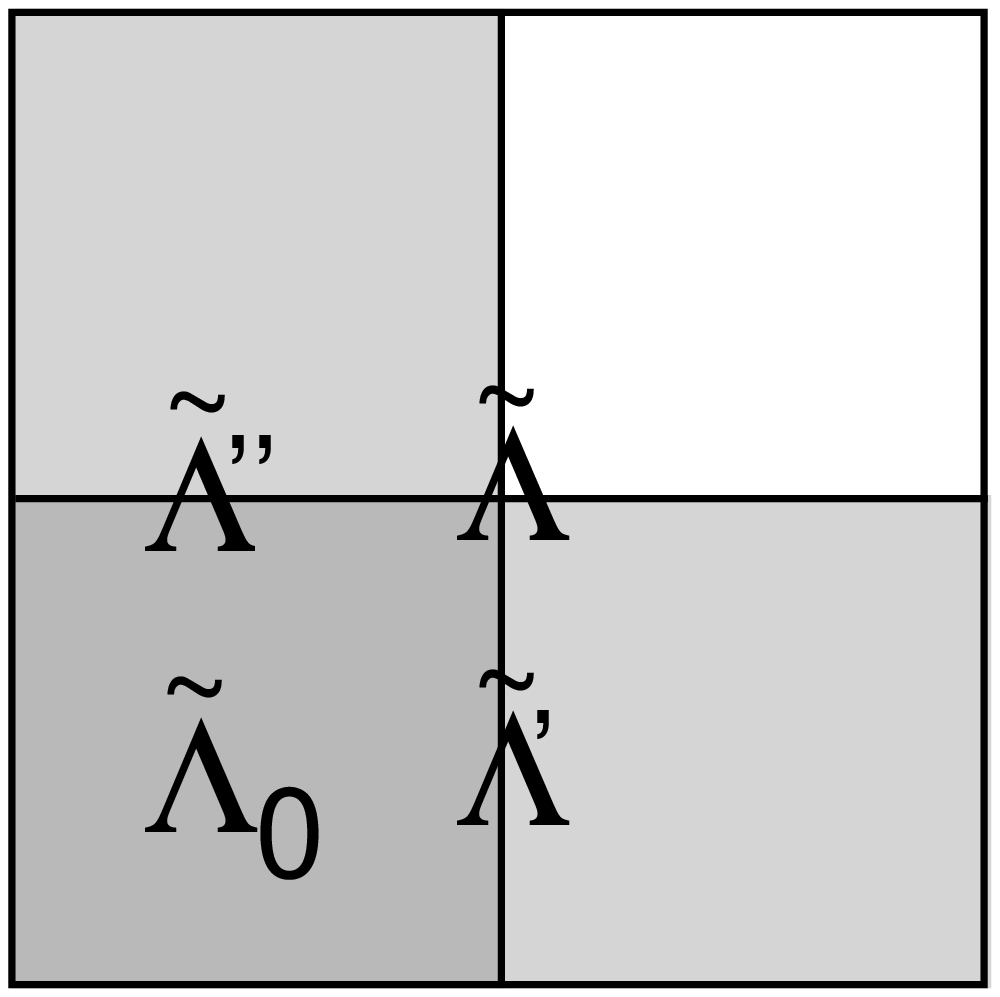}
\caption{We present a cross section through the 0-2 plane of the various 
lattices that appear in the Nahm transformation with twisted boundary
conditions $n_{02}=-n_{20}=1$ on $R^4/\Lambda$. Units are such that 
$l_0^2=l_2^2=\half$, for which $R^4/\tilde\Lambda_0$ is of the same 
size as the original torus. In this case $N_0=2$ and the two choices for 
the minimal lattices with twist are denoted by $\Lambda'$ and $\Lambda"$. 
The dual of the lattice $\Lambda_0$ gives the torus on which the Nahm 
transformed gauge field lives, with the same twist. Duplications on the 
dual side are identical, except for interchanging the roles of $\Lambda$ 
and $\Lambda_0$ (and less importantly $\Lambda'$ and $\Lambda"$). When a 
symbol overlaps with different cells it belongs to all of those cells 
as a whole.}
\end{figure}

Each choice of $\Lambda'$ will lead to a different set of zero-modes. This
can only mean that the resulting dual gauge fields are related by gauge
transformation. Gauge invariant quantities have to agree when derived from
the various choices of $\Lambda'$. But this means that the smallest unit
cell in terms of which these gauge invariant quantities can be reproduced,
is the intersection of the unit cells of all $\tilde\Lambda'$, which is 
precisely $\hat\Lambda=\tilde\Lambda_0$, the dual of $\Lambda_0$ (in the 
same way the unit cell of $\Lambda$ is the intersection of the unit cells 
of all $\Lambda'$). Note that $\Lambda_0$ is the smallest sublattice of 
$\Lambda$, that is also a sublattice of all possible choices of $\Lambda'$ 
(the unit cell of $\Lambda_0$ consists of $N_0$ unit cells of $\Lambda'$ 
and therefore of $N_0^2$ unit cells of $\Lambda$). With the twist in the standard form given 
above, $\Lambda_0$ is obtained by duplicating the hypercube that defines the 
original torus $R^4/\Lambda$, $(p_1-1)$ times in {\em both} the 0 and 2 
directions and $(p_2-1)$ times in {\em both} the 1 and 3 directions. 
In Fig.~1 we elucidate this for the 0-2 plane with $q_1=1$ and $p_1=2$.

This shows that  $\hat A$ can be defined as an $SU(N_0Q)$ gauge field
on $R^4/\tilde\Lambda_0$~\cite{Ntbc} and has topological charge $N/N_0$.
Nevertheless, as is obvious in particular for $N/N_0$ not an integer, the
unit cell of $\Lambda_0$ cannot be without twist, i.e.  $\hat A$ is also a
gauge field with {\em twisted boundary conditions}. For $\SUT$ in the case that the 
original twist is already in the Frobenius standard form (where the situation 
of Fig.~1 applies in the 0-2 and/or the 1-3 planes), the dual twist can be 
read off from the minimal number of duplications required to 
go from $\tilde\Lambda_0$ to any of the choices $\tilde\Lambda'$. 

\section{The case of $SU(2)$ (twisted) instantons}

Having presented the general formalism, let us now concentrate on the study
of $\SUT$  gauge field configurations for spatially symmetric tori. From the 
previous formulas one can deduce that the symmetric structure of the torus is 
not always preserved by the Nahm transform. Nonetheless, the dual torus is 
always contained within the symmetric torus $R^4/\tilde{\Lambda}$. Furthermore,
if a given torus has $l_s/l_t\gg 1$, its Nahm transform 
corresponds to $l_s/l_t\ll 1$. Thus, the Nahm transform of the 
torus instanton configurations are the torus caloron configurations. The space 
of gauge inequivalent solutions defines the moduli space. Its dimension is 
given through an index theorem to be $4NQ$ for $\SUN$ gauge fields of charge 
$Q$ on the torus. Note that $QN=\hat{Q}\hat{N}$ so that the dimensionality is 
preserved by the Nahm transform. Indeed, in Ref.~\cite{BrvB} it is proven that 
the Nahm transformation induces an isometry between the moduli spaces. 

\subsection{The twisted instanton $Q=\half$}

We start with the smallest value of the topological charge, for $SU(2)$
this is $Q=\half$. The dimension of the moduli space is 4, so that the set 
of solutions is discrete up to translations. The value of the topological 
charge is related to the twist by the formula~\cite{Tho1,Pie2}
\be
Q=\nu-\frac{\vec k\cdot\vec m}{N}\quad,
\ee
where $\nu$ is an integer. Fractional values of the topological charge are
attainable for non-orthogonal twists, i.e. $\vec k\cdot\vec m\ne 0\bmod N$.
For $\SUT$ we choose $\vec k=\vec m=(0,1,0)$, with $\vec k\cdot\vec m=1$.
We deduce $N_0=4$ and therefore the Nahm transformation is again an $\SUT$
self-dual gauge field with topological charge $\half$. Furthermore, it
lives on a torus with  generators $\hat{e}^{(\mu)}=\half\tilde{e}^{(\mu)}$,
where $\tilde{e}^{(\mu)}$ are the dual basis vectors $\tilde e^{(0)}=
(1/l_t,0,0,0)$,  $\tilde e^{(1)}=(0,1/l_s,0,0)$, $\tilde
e^{(2)}=(0,0,1/l_s,0)$, $\tilde e^{(3)}=(0,0,0,1/l_s)$.
The corresponding twist is unchanged.

Now we will explain how this works for configurations with $l_s/l_t\ll1$. 
This will serve to illustrate the precision of our numerical techniques. One  
can generate self-dual configurations numerically by employing the method 
described in Ref.~\cite{MTB}. To minimise the errors due to lattice artifacts 
we actually make use of the improved cooling technique described in 
Ref.~\cite{MaSn}, with the parameter $\eps=0$. One can implement twist 
by the method of Ref.~\cite{Gro}. We have generated self-dual configurations 
with the twist mentioned above on lattices of size $4 \times 16^3$ and 
$32 \times 8^3$. Notice that in the former case $l_s/l_t=4$ and in the latter 
case $\frac{1}{4}$. They correspond to Nahm-dual tori for this twist. In both 
cases the action density of the configuration has a lumpy profile with a 
single maximum in space. The position of the maximum does not coincide with 
a lattice point, but can be estimated by interpolation.  
Then, given the
interpolated position of the maximum, one can 
determine the coordinates corresponding to each lattice point with respect to it.  

\begin{figure}[htb]
\vspace{9.5cm}
\includegraphics{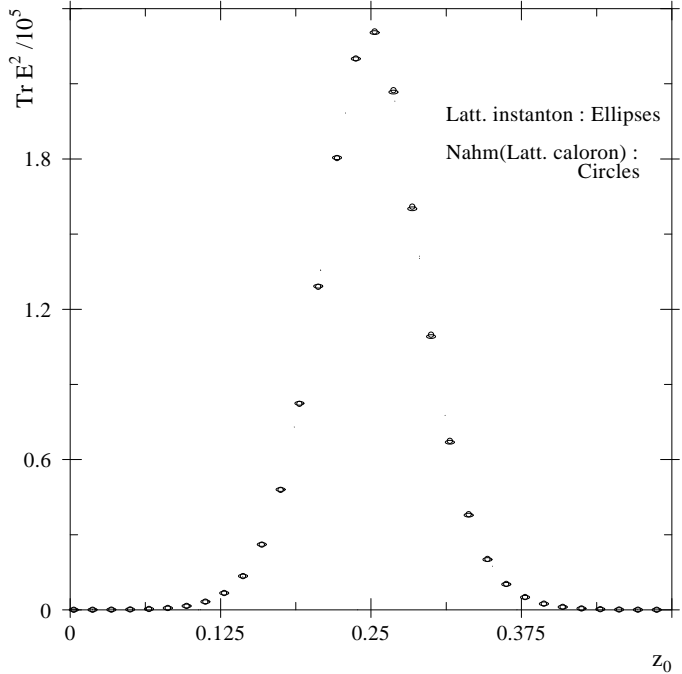}
\caption{Comparison between the action densities of a numerical $Q=\half$ 
instanton with twist $\vec{k}=\vec{m}=(0,1,0)$, living on a $32 \times 8^3$ 
lattice, and the Nahm transformation of a $Q=\half$ caloron (BPS 
monopole) with the same twist, living on a $4 \times 16^3$ lattice. The 
depicted values correspond to points in the $32 \times 8^3$ lattice, along the 
lattice direction $\vec{n}_{latt}=(8,4,8)$; the peak of the action density 
of the lattice instanton being interpolated, in lattice coordinates, to $n=(15.809,3.741,3.979,4.162)$
and shifted to $(\quart,\sixteenth,{\scriptstyle 0},\sixteenth)$ where the maximum
of the Nahm transform of the caloron sits. 
The continuum time period has been set to $l_t=1$ for the caloron
and hence $\tilde{l_t}=\half$ for its Nahm transformed instanton.}
\end{figure}

What we did next was to apply the numerical Nahm transformation introduced in
Ref.~\cite{Nlat}. The values of $z$, where the Nahm transformation was 
calculated, were those corresponding to the lattice points of  the dual 
configuration. In this way the Nahm 
transformation can be compared with the lattice configuration obtained on 
the dual torus. Fig.~2 shows the result of such a comparison for 
the action density along a line on the torus. The agreement is remarkable. It 
shows the way in which our expectations work and gives us confidence on the 
precision of our numerical methods to investigate other more complicated 
situations. Note that in fact Ref.~\cite{Nlat} dealt with the case of four 
$Q=\half$ instantons (with $l_s/l_t=1$) on a $12\times6\times12\times6$ lattice,
finding that the configuration is mapped to itself (up to a shift). In terms 
of the $Q=\half$ twisted instantons this can be understood from the fact that 
only its position is a free parameter. Since the work in Ref.~\cite{Nlat} 
the numerical methods have been greatly improved, details of which will be 
reported elsewhere. 

\subsection{Temporal twist $Q=1$}

Let us recall what is known about  $Q=1$ configurations with $\vec m=\vec 0$
and $\vec k\neq\vec 0$, for the cases $l_t/l_s\rightarrow 0$ (torus 
calorons) and $\infty$ (torus instantons). These will be mapped onto each
other by the Nahm transformation and we are interested in establishing how
the 8 dimensional moduli spaces are related.

The torus caloron configurations ($l_t\ll l_s$) have been studied recently by 
$\eps$-cooling techniques~\cite{Calo}. As $l_t/l_s\rightarrow 0$ these 
configurations approach some of the non-trivial holonomy 
calorons~\cite{ThPi,Lee}, namely those associated with $\omega=\frac{1}{4}$. 
The parameter $\omega=|\vec\omega|$ is defined through the holonomy that in 
the infinite volume is defined by the value of the Polyakov loop at infinity
\be
\label{tihol}
\exp(2\pi i\vec\omega\, \cdot\vec\tau)=\lim_{|\vec x|\rightarrow\infty}
\Pexp(-i\int_0^{l_t} A_0(t,\vec x)dt)\quad.
\ee
These calorons with $\omega=\quart$ correspond to equal size constituent 
monopoles located at two different points. For finite $l_s/l_t$ the periodic
boundary conditions in space slightly modify  the action density profiles 
without changing the qualitative features. Already for $l_s/l_t=4$ the
agreement is excellent~\cite{Calo}. 

\begin{figure}[htb]
\vspace{5.5cm}
\includegraphics{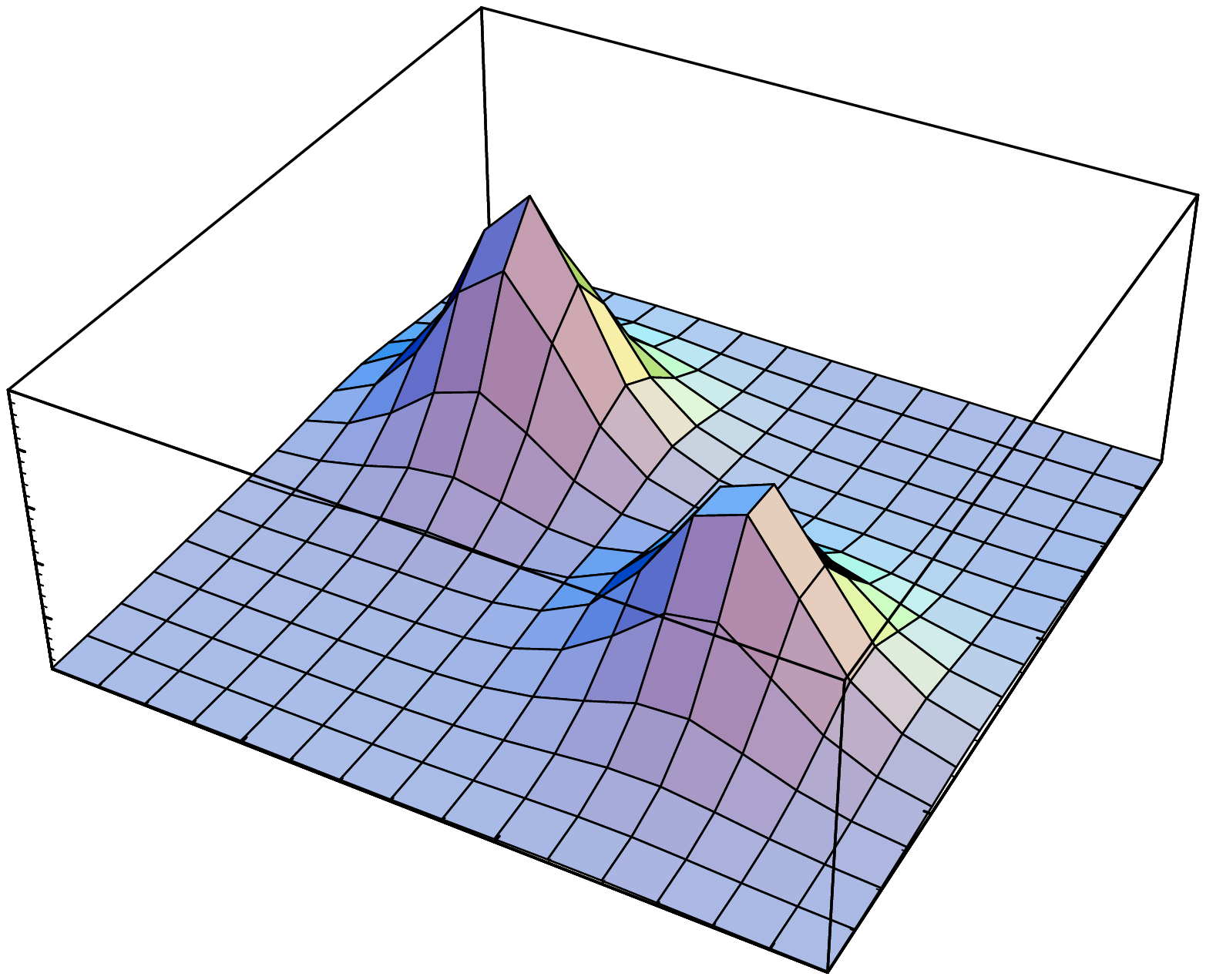}
\includegraphics{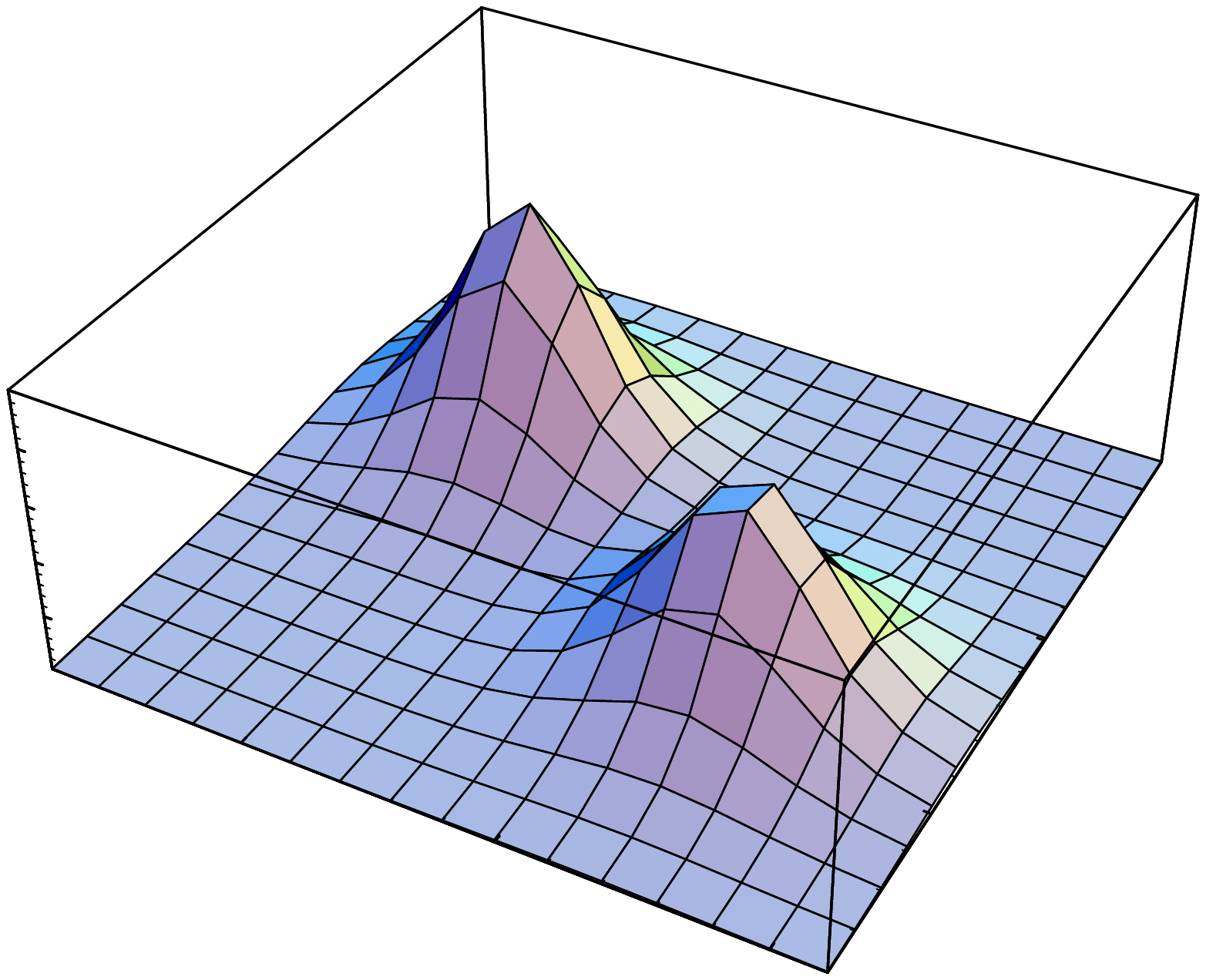}
\caption{Zero-mode density profiles for the two zero-modes of the lattice 
caloron (left) on a $4\times 16^3$ lattice for $\vec k=(1,1,1)$, created with 
improved cooling ($\eps=0$). The profiles fit well to the two zero-modes for 
the infinite volume analytic caloron solution (shown on the right at 
$y\!=\!t\!=\!0$) with $\omega=\quart$ and constituents at 
$\vec y_1=(2.50,0.12,0.95)$ and $\vec y_2=(1.38,-0.24,2.67)$, in units where 
$\beta=l_t=1$ (or $a=\quart$) and the left most lattice point corresponding to 
$x=z=0$. The plots give the added densities of the two zero-modes.}
\end{figure}

Recently also the Weyl-Dirac zero-mode for the infinite volume caloron with 
non-trivial holonomy was determined analytically~\cite{MTCP}. The zero-modes 
are more localised than the action density and the agreement between the 
numerically determined zero-mode for the torus caloron is therefore expected 
to be even better. This is illustrated in Fig.~3, where we compare 
these exact infinite volume zero-modes with those that are numerically 
constructed. Since the zero-modes are the ingredients for performing the Nahm
transformation, the present comparison shows the numerical accuracy that was 
achieved. For convenience we have taken for this comparison the parameters 
that were used in Fig.~1 of Ref.~\cite{Calo}, where one can find the action
density profiles. In this particular case $\vec k=(1,1,1)$ and we duplicate
the initial torus  in the time-direction, to reach a situation without twist.
This gives a charge 2 configuration with two zero-modes, of which one is 
periodic and the other is anti-periodic. From Ref.~\cite{MTCP} one knows that
each is localised on one of the constituents. The numerical procedure was to 
combine the two zero-modes of the doubled lattice in two orthonormal modes, each
maximising the overlap with one of the two constituents. Using the large degree
of localisation, we have simply added the zero-mode densities together to 
show the result in one figure, but we stress each lump corresponds to the 
density of one of the zero-modes. There are no free parameters involved in 
this comparison and the agreement for the zero-modes is indeed impressive. 

In the infinite volume the parameters of the caloron are described by the 
spatial positions of the two constituents (6 parameters), one overall time 
position and a $U(1)$ phase rotation that can be undone by a gauge 
transformation. For large separations of the constituents the caloron becomes 
static with the two constituent monopoles showing up as separate lumps. On 
the other extreme when the constituent monopoles are close to each other they 
fuse into a single lump that looks like an ordinary instanton in $R^4$.

Torus instanton configurations ($l_s\ll l_t$) were studied a few years 
ago~\cite{MaSn,MaPi}. As $l_s/l_t\rightarrow 0$ the configurations evolve 
into well-defined solutions on $T^3\times R$. One feature shared by all these 
torus instanton solutions is that they are local in time. At large times 
the configurations have to approach a pure gauge, to guarantee the action 
stays finite. The gauge fields with zero curvature on $T^3$ at $t=\pm\infty$ 
are characterised by their holonomies (spatial Polyakov loops)
\be
\label{sphol}
P_j^\pm=\lim_{t\rightarrow\pm\infty}\Pexp(-i\int_0^{l_j}A_j(t,\vec x)dx_j)\quad.
\ee
In the presence of a temporal twist $\vec k$, the holonomies at both ends 
are related
\be
\label{pol}
P_j^+=e^{2\pi i k_j/N}\,P_j^-\quad.
\ee
When we consider $l_t/l_s$ finite, the instanton has 8 parameters, but the 
holonomies can not be rigorously defined. In the limit $l_t/l_s\rightarrow
\infty$ they would, however, be related by the twist factors. It is likely 
that in this limit the three holonomies are part of the moduli that describe 
the solution (for the infinite volume caloron, gauge zero-modes associated 
with a variation of $\omega$ are not normalisable and $\omega$ is {\em not} 
a moduli parameter). There are indications
that the holonomies fix the scale of the torus instanton~\cite{MaPi,Pie3}. In 
addition there are of course the four position parameters, and presumably a 
phase that describes the colour orientation relative to the twist matrices.

The dual gauge field for instantons on $T^3\times R$ with temporal twist
was shown to be a solution of the abelian Bogomolny equations with point
sources~\cite{Pie3} that have (due to the self-duality) equal magnetic and
electric charges. From this one computes the Maxwell field analytically by
summing over the periodic copies of these charges. Essential is that the
location of the point charges is determined by the holonomies $P_j^\pm$.
At these specific points on the dual torus the Nahm transformation is
modified due to boundary terms in the non-compact directions. This leads
to four singularities, whose locations are determined by requiring
$\exp(-2\pi i l_j z_j)P_j^\pm$ to have a unit eigenvalue (this leads to a
zero-energy state in the Weyl-Dirac Hamiltonian at $t=\pm\infty$ and
therefore to non-exponential decay of the zero-mode). The charges are all
equal in absolute value, due to Dirac quantisation~\cite{Pie3}, and opposite
in sign for those associated to the holonomies at either end. With temporal
twist, the location of the sources with one sign of the charge are displaced
by   $\half (\frac{k_1}{l_1}, \frac{k_2}{l_2} ,\frac{k_3}{l_3})$
 with respect to the location of the  two sources with the opposite sign.

\begin{figure}[htb]
\vspace{6.5cm}
\includegraphics{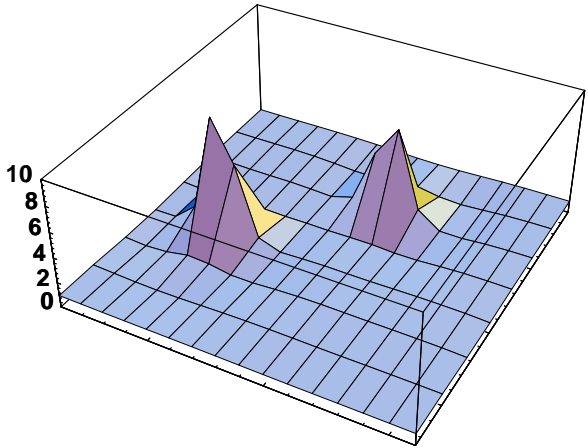}
\caption{Ratio of the action densities of the Nahm transformed gauge field
for a $l_t \times 8^3$ lattice, dividing the $l_t=40$ result by that of 
$l_t=20$. The (approximate) holonomies are given by $\half\Tr P_j^-=-\half
\Tr P_j^+=(0.86,0.76,0.082)$.}
\end{figure}

We illustrate in Fig.~4 how the singularities arise as $l_t$
increases. The results are based on a charge 1 instanton on an $20 \times 8^3$
and $40\times 8^3$ lattice with temporal twist $\vec k=(1,1,1)$ and holonomies
defined by $\half\Tr P_j^-=-\half\Tr P_j^+=(0.86,0.76,0.082)$ (changing
$l_t$ and making sure the holonomies stay fixed can be achieved by cooling
with open boundary conditions, introduced in Ref.~\cite{MaPi}). We choose
units such that $l_s=1$, so that  the sources with positive charge are located at
$\vec z=(0.415,0.387,0.263)$ and $\vec z=(0.585,0.613,0.737)$ and the ones
with negative
charge at $\vec z=(0.915,0.887,0.763)$ and $\vec z=(0.085,0.113,0.237)$.
Fig.~4 is plotted in the plane defined by $z_3=0.263$, which
is through the position of the first source and still close to the position
of the last source. Shown is the ratio of the action density for the
$40\times 8^3$ lattice over the action density for the $20\times 8^3$ lattice.

The nature of the singularities is illuminated by the observation, made
before, that the Nahm transformation of the  torus instantons
are  the torus calorons. Thus, the action density peaks correspond to the
location of the BPS constituent monopoles. Their mass (associated to
$w=\quart$) should be given by $8 \pi^2 l_t$ (in units where $l_s=1$). The result is compatible
with the fact that the peak heights of these BPS monopoles scale as $l_t^3$.
Away from the core of these BPS monopoles the field becomes abelian and
should become independent of $l_t$, as the charge is fixed. Fig.~4
shows that indeed the ratio becomes one outside of the core of the monopoles.
The non-abelian core of the BPS monopole shrinks to a point in the limit
$l_t \rightarrow \infty$, and one is left with the resulting abelian field.
This is illustrated in Fig.~5, where  we plot the square of the electric field
in the plane defined by $z_3=0.5$. The numerical result is based on the
$40\times 8^3$ lattice, for which the plane $z_3=0.5$ is far enough away
from all sources not to be affected by their non-abelian dressing 
as $\SUT$ BPS monopoles of mass $40\pi^2$ (in units where $l_s=1$).
The agreement, that involves no free parameters, is indeed very good.

\begin{figure}[htb]
\vspace{5.4cm}
\includegraphics{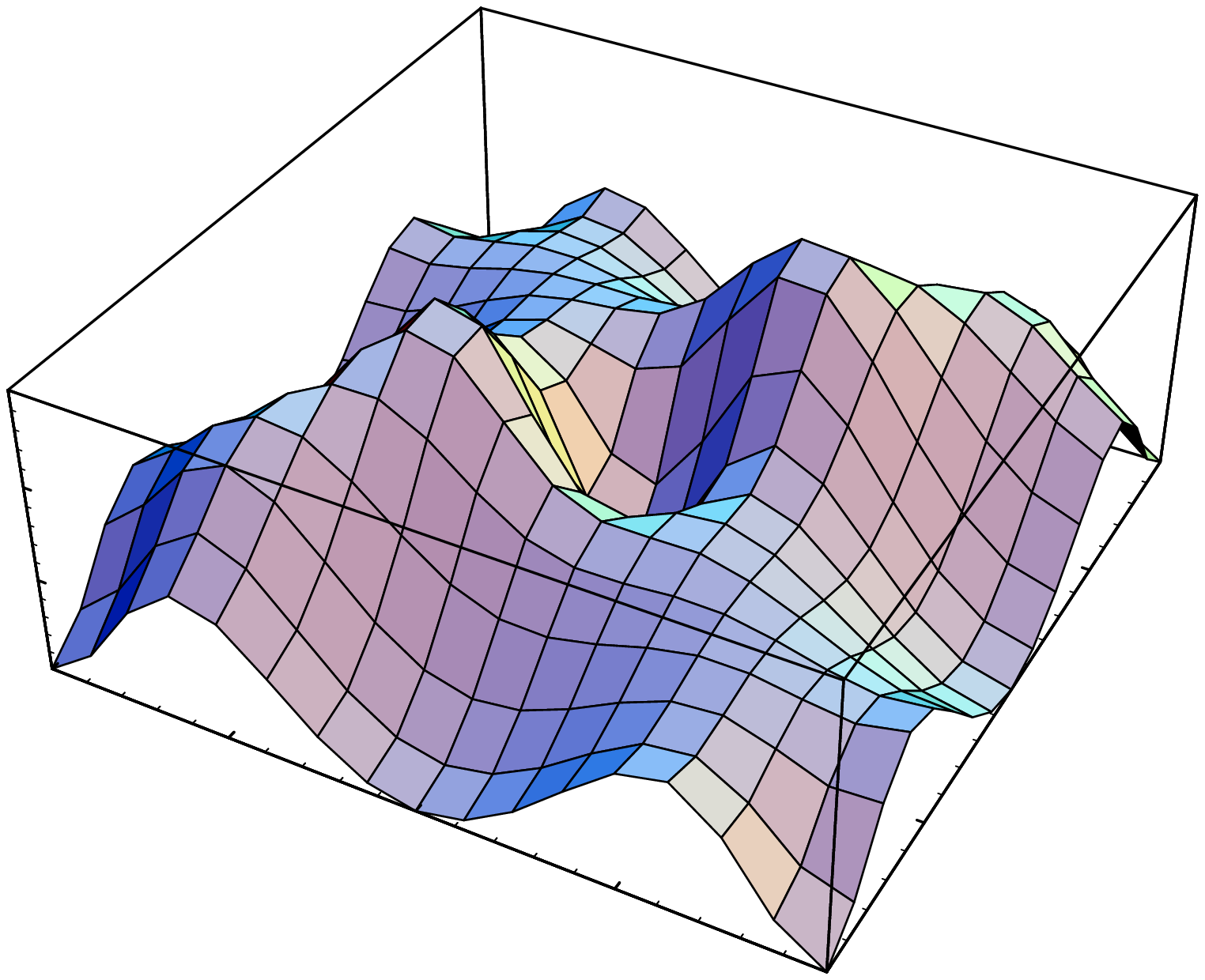}
\includegraphics{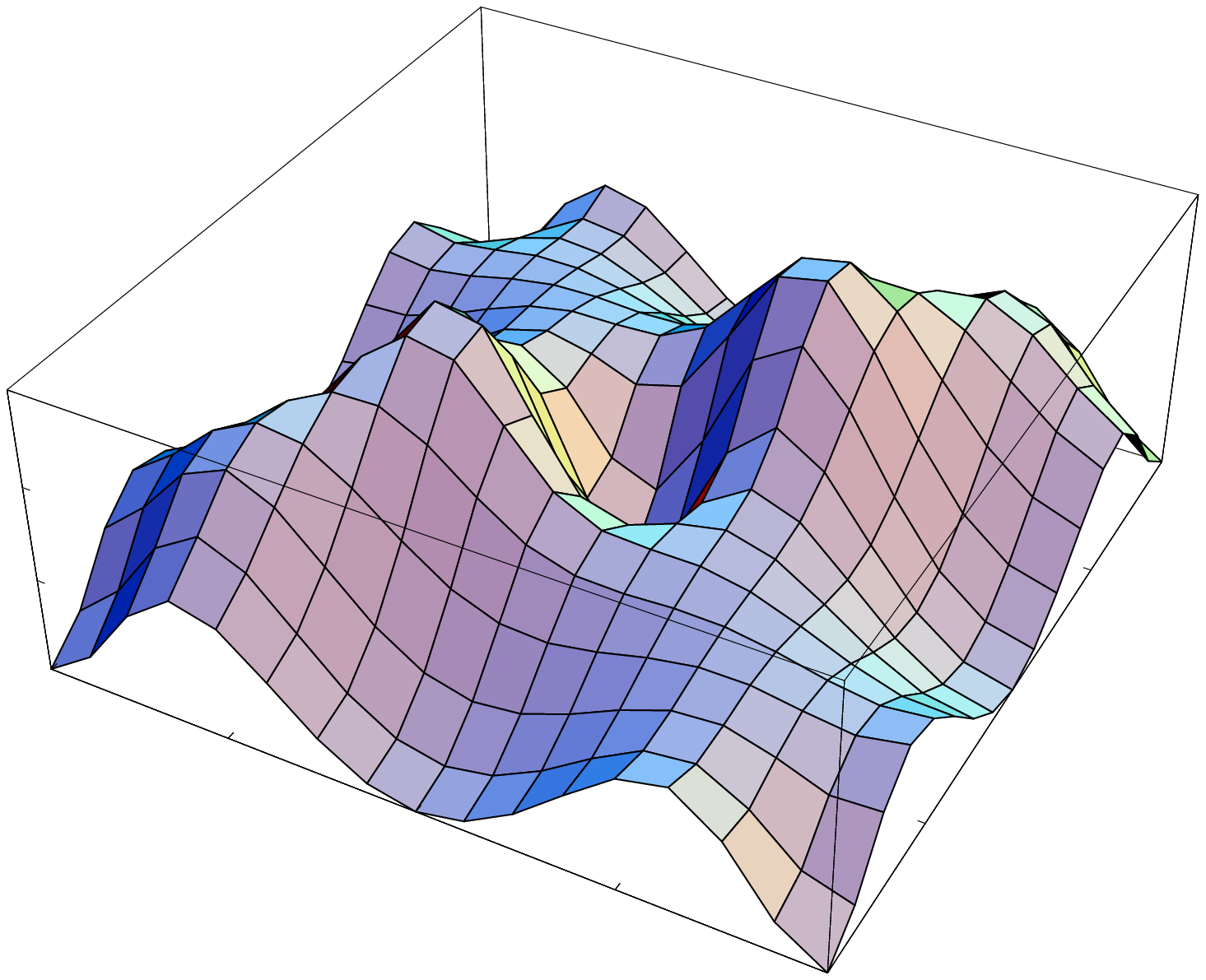}
\caption{Comparison of $\vec E^2$ in the plane $z_3=0.5$ for the Nahm
transformation of the $40\times 8^3$ instanton configuration described in
the caption of Fig.~4. On the left is the numerical result and on the right 
the analytic result for $T^3\times R$.}
\end{figure}

The Nahm transformation was constructed by doubling in time, such that at 
finite $l_t$ this corresponds to a charge 2 torus instanton, which is mapped 
to a charge 2 torus caloron defined on the hypercube $[0,\half l_t^{-1}]
\times[0,1]^3$. Our general formalism tells us that this is formed by duplicating 
the Nahm-dual torus, which is half the size. The latter corresponds to
a skew lattice, since the twist is not in the Frobenius standard form.
The additional lattice generator  is precisely  the vector $\half
(0,\frac{k_1}{l_1}, \frac{k_2}{l_2} ,\frac{k_3}{l_3})$ giving the displacement 
of the 2 positive versus the 2 negative sources.

In summary, we   have illustrated how  the Nahm transformation maps torus 
instantons into  torus calorons, and how the (approximate) holonomies are
mapped into the 
positions of the constituent BPS monopoles. This relation is exact in the
limit that $l_t\rightarrow \infty$ (resp. 0). To test its  validity 
 at finite $l_t/l_s$ we studied  the case with temporal twist $\vec k=(0,1,0)$
($\vec m=(0,0,0)$). One cannot have both the original and Nahm-dual torus to be 
spatially symmetric, due to the asymmetric twist, but that is not important
for the general conclusion  about the mapping we are studying. 
If our original torus has size $l_t\times l_s^3$, $\hat{\La}= \tilde\Lambda_0$ has size 
$\frac{1}{2 l_t}\times\frac{1}{l_s}\times\frac{1}{2l_s}\times\frac{1}{l_s}$. 
The twist vectors for the Nahm transform are the same as the original ones. 

By means of improved cooling ($\eps=0$) we have generated a lattice torus caloron 
configuration~\cite{Calo}. The relative distance between the two constituent 
monopoles can be tuned by using different values of the parameter
$\eps$~\cite{Calo}. On 
the other hand, it is possible to use a modified version of cooling~\cite{MaPi} 
(fixing the field to have zero colour magnetic fields at $t=0$ and $t=l_t$ 
in terms of prescribed holonomies) in order to produce torus instanton 
configurations with fixed holonomies. If we choose these to be the ones 
determined by the location of the monopole constituents of the torus caloron,
then the numerical Nahm
transformation~\cite{Nlat} can be applied to each of these configurations
and has to give back the other configuration, up to a translation. This is illustrated in Fig.~6.
The somewhat larger numerical differences are in part due to
the not sufficiently large ratio for $l_t/l_s$ which, for the required values of the
holonomies, enforces a rather narrow instanton which is not easily stabilised
on the lattice. Results for a smaller value of $l_t/l_s$ (not shown) 
confirm this improvement with increasing $l_t/l_s$. Going to even a larger
ratio is however computationally too expensive.
The comparison confirms that the holonomies of the time-twisted torus instanton are part
of the moduli.
 
\begin{figure}[htb]
\vspace{9cm}
\includegraphics{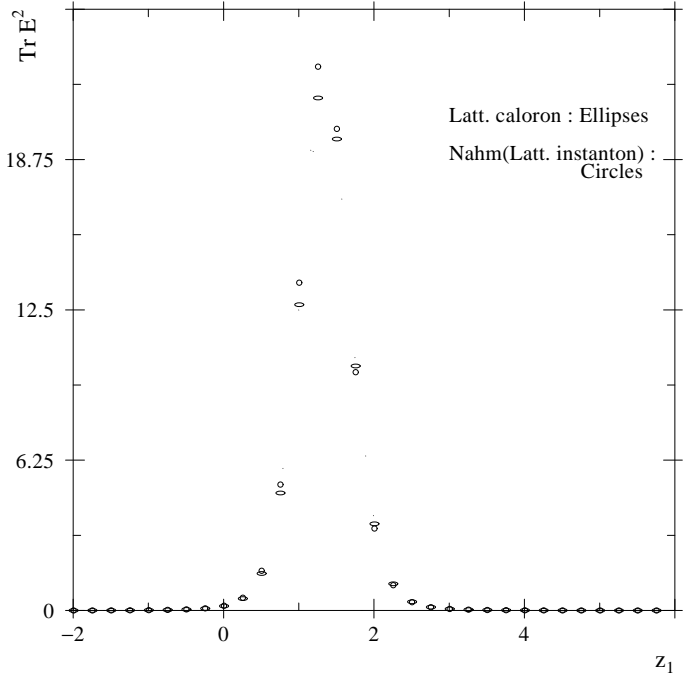}
\caption{Comparison between the action densities of a numerical $Q=1$
instanton with twist $\vec{k}=(0,1,0)$ and $\vec m=\vec 0$, living on a 
$32\times8\times4\times8$ lattice, and the Nahm transformation of $Q=1$ 
torus caloron with the same twist, living on a $4\times32^3$ lattice. The 
constituent locations were used to fix the holonomies in generating the torus 
instanton.  Depicted values correspond to points in the $4\times32^3
$ lattice, along the $x$ direction. The 
continuum units are such that $l_t=1$ for the caloron.}
\end{figure}

\subsection{Spatial twist $Q=1$}

We will conclude by discussing the case of spatial twist ($\vec k=\vec0$), for 
which we take $\vec m=(0,1,0)$. If our original torus has  size $l_t\times 
l_s^3$, the Nahm dual torus for this twist has size $\frac{1}{l_t}\times
\frac{1}{2l_s}\times\frac{1}{l_s}\times\frac{1}{2 l_s}$ (to obtain instead a 
spatially symmetric dual torus we should apply the Nahm transformation to a 
torus with size $l_t\times\half l_s\times l_s\times\half l_s$). As before, 
the dual gauge field has the same spatial twist as the original gauge field. 
The phenomenology of these configurations, for the large aspect ratios we
are interested in, is different to the previous case with temporal twist.
But there are also quite interesting analogies that may be important for the
case $l_t=l_s$, where from the Euclidean point of view there should be no
essential difference between twist in space or time.

The analysis of Ref.~\cite{Calo} shows that for $l_s/l_t$ large, torus calorons
with spatial twist are given by configurations that approximate the infinite 
volume caloron with arbitrary holonomy, parametrised by $\omega\in[0,\half]$, 
such that the two constituent monopoles have in general differing masses, 
$16\pi^2\omega/l_t$ and $8\pi^2(1-2\omega)/l_t$. But now their relative 
positions are fixed and determined by the spatial twist. This contrasts with 
the situation for temporal twists for which the value of $w$ is fixed, but 
the relative position is arbitrary.

Torus instantons with twisted boundary conditions in space, and $l_s/l_t$ 
small, have been studied extensively in relation to the Hamiltonian
formulation in a twisted box~\cite{RTNcol}. The outcome of these
studies~\cite{TPA} showed that the configurations can be described in 
terms of two $Q=\half$ twisted instantons ($2Q$ twisted instantons in the case
of higher charges), that have both twist in space and time. The net twist
in time, however, vanishes.  We can view the $Q=\half$ constituents of the 
$Q=1$ instanton in very much the same way as the monopole constituents for 
the caloron. It was found that the space-time locations of these twisted instanton 
lumps can be arbitrary. Only when they are very close to each other, they 
merge into a single lump. In that region, when the size of the lump becomes 
small compared to $l_s$, the configuration behaves like the ordinary 
instanton in $R^4$. There are two very simple specific examples where this 
constituent nature of the torus instantons with spatial twist is apparent.
Take a $Q=\half$ instanton with twist $\vec{m}=\vec{k}=(0,1,0)$ and duplicate 
either in the 0 or 2 direction to remove the time twist. The result is a 
$Q=1$ instanton with spatial twist. Either the two lumps have the
same spatial positions and are separated maximally along the time direction,
or they have equal time positions and are maximally separated along the space
direction.

We now turn our attention to studying how the Nahm transformation relates 
these torus instantons and calorons. For the above example built from two 
$Q=\half$ instantons, this can directly be understood in terms of the
results of section 3.1, where it was shown that the dual of such a $Q=\half$
instanton is a BPS monopole, with twist $\vec k=\vec m=(0,1,0)$. Duplication
of the instanton in the 0 direction is dual to duplication of the BPS 
monopole in the 2 direction, and we do get a caloron with maximal separation
of the ($\omega=\quart$) constituents in the direction of the spatial twist. 
On the other hand, duplication of the instanton in the 2 direction is dual to 
duplication of the BPS monopole in the 0 direction. This doubles the mass of 
the constituent monopole and corresponds to a caloron with trivial holonomy 
$\omega=0$ or $\half$ (the other constituent is massless and therefore absent). 
This establishes that, with twist in space, $\omega$ need not be $\quart$. 

To deal with a more  general case,  we consider the $Q=1$ instanton with spatial 
twist being made out of two $Q=\half$ ($\vec{m}=(0,1,0)$, $\vec{k}$ arbitrary 
such that $\vec k \cdot\vec m=1 ({\rm mod}\,2$)) with an arbitrary time separation, i.e.
no longer obtained by duplication in the time direction. To study some of the 
properties of its Nahm transform, we consider the limit $l_s\rightarrow 0$. 
In this limit all spatial dependence is removed and the two $Q=\half$
instantons shrink to two points, specified by their positions in time. In 
one dimension curvature can be non-trivial only at point-like singularities, 
so the curvature vanishes everywhere, except at the locations of the $Q=\half$
instantons, where the field strength becomes singular. The curvature free 
regions are described by a constant abelian gauge field $\hat A_j$, 
making a jump at the positions of the $Q=\half$ instantons. In this limit, for 
a single $Q=\half$ instanton, the twist determines the jump in the constant 
gauge field to be $\Delta\hat A_j=\pi k_j/l_s$ (relevant for the case of space 
twist, so as to assure that Eq.~(\ref{pol}) is satisfied. In the general case,
where the values of the Polyakov loops are not fixed (see below), 
there is more freedom). The $Q=1$ instanton is therefore described by 
\be
\label{abel}
\hat A_j=2\pi(\xi_j+\half k_j\chi_{[t_a,t_b]}(t)/l_s)\quad,
\ee
where $\xi_j$ (and $\hat A_0$) are arbitrary constants (fixed for spatial 
twist, see later) and $\chi_{[t_a,t_b]}(t)$ is the characteristic function 
of the interval $t\in[t_a,t_b]$. This
result agrees exactly with  the Nahm transformed gauge field of the infinite 
volume caloron~\cite{ThPi}. We can identify $\half\vec k/l_s(=\hat a\pi\rho^2)$
with the distance vector between the two constituents (in units set by $l_t=1$) 
and $\omega=\half(t_b-t_a)$ with its holonomy (see the second of 
Ref.~\cite{ThPi}, Eqs.~(41) and (65), for the precise definitions of $\rho$ 
and $\hat a$). 

\begin{figure}[htb]
\vspace{9cm}
\includegraphics{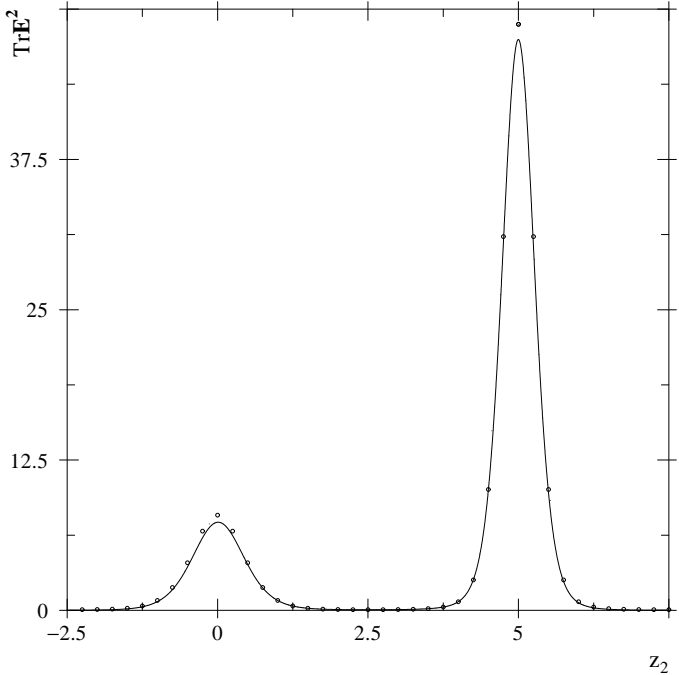}
\caption{Comparison between the action densities of the numerical Nahm
transform of a torus instanton and the appropriate infinite volume analytic 
caloron solution. The original field lives on a $80\times 4\times 8\times 4$ 
lattice with twist $\vec{k}=(0,0,0)$, $\vec{m}=(0,1,0)$. Units are set by
$l_t=1$. The temporal distance between the two $Q=\half$ constituents for 
the instanton corresponds to a caloron with $\omega=d_0/2=0.311$. The plot 
is along the line connecting the centers of the two constituent monopoles, 
defined by $z_1=z_3=5$ as determined by the spatial holonomies of the 
instanton.}
\end{figure}

It may seem that for the Nahm gauge field of the caloron $t_a=-\omega$ and 
$t_b=\omega$ are fixed, but one should of course realise that $\hat A$ 
is still self-dual, away from the singularities, when $t$ is shifted over a constant.
We conclude that the holonomy of the caloron 
fixes the relative locations of the $Q=\half$ instanton lumps under the 
Nahm transformation, whereas the holonomies of the instanton, determined 
in each of the two flat regions (for $l_s\rightarrow 0$), fix the relative locations 
of the constituent monopoles of the caloron. Their difference vector 
$\half\vec m/l_s$, remarkably, agrees with maximal separation on the dual 
torus along the direction determined by the spatial twist\footnote{Note that 
on the dual torus a displacement by $\frac{1}{2l_s}$
along directions 1 and 3 corresponds to a shift by a full torus period.
Hence, a shift by $\half\vec k/l_s$ is identical to $\half\vec m/l_s$, for
any $\vec k$ with $\vec k\cdot\vec m=1({\rm mod}\,2)$.}.
 This derivation is as rigorous, in the limit $l_s\rightarrow 0$, as 
the one for $T^3\times[0,l_t]$, in the limit $l_t\rightarrow\infty$ and has 
given a beautifully consistent picture of the dual relationship between the 
moduli spaces of the calorons and finite volume instantons.

To test this picture at finite $l_s$ we have made use of our numerical methods.
We started with a configuration on an $80\times4\times8\times4$ lattice
containing two $Q=\half$ torus twisted instantons (with $\vec m=\vec k=(0,1,0)$)
separated in time by
a distance $d_0$. In units where $l_t=1$ (so $l_s=1/10$), the
lattice $\Lambda$ is generated by $e_0=(1,0,0,0)$, $e_1=(0,l_s/2,0,0)$,
$e_2=(0,0,l_s,0)$ and $e_3=(0,0,0,l_s/2)$. The dual torus  $R^4/\tilde\Lambda_0$
has as unit cell $[0,1]\times[0,l_s^{-1}]^3$. This configuration is constructed in
several steps. First one starts with a $Q=\half$ configuration on a
$40\times4\times8\times4$ lattice with twist $\vec{k}=\vec{m}=(0,1,0)$.
Using a time reversal and parity transformation one obtains a second $Q=\half$
instanton with the Polyakov loops at $t\approx\pm\infty$ interchanged.
One can glue the two configurations together at arbitrary time separation
$d_0$ and transform them appropriately to trivial twist matrices
in time. If $d_0\gg l_s$ this procedure gives an approximate self-dual solution 
on the  required torus and with spatial twist. The small deviations from 
self-duality induced by the procedure can be eliminated  by cooling with $\eps=0$.
We end up with  a $Q=1$ configuration with the desired properties.

\begin{figure}[htb]
\vspace{9cm}
\includegraphics{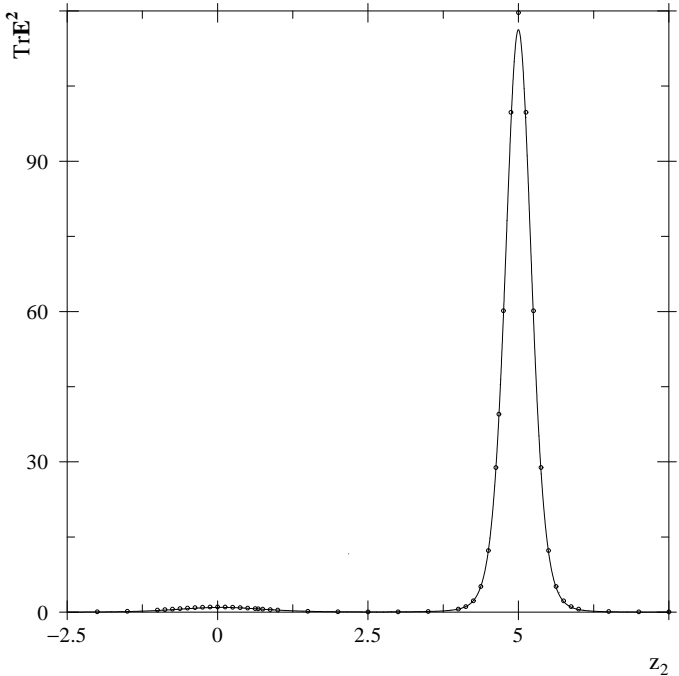}
\caption{The same as in Fig.~7, but for a temporal distance corresponding to 
$d_0=0.786$.}
\end{figure}

A few remarks are in order, concerning the values of the Polyakov loops
for the space twisted instantons. Eqs.~(\ref{tihol}) and (\ref{sphol}) are
only valid in the gauge where the appropriate $\Omega_a(x)$ are trivial.
Such a gauge was assumed for the discussion in sect.~3.2. The general
formula for a Wilson loop that closes by the shift over a period
reads~\cite{Pie1}
\be
\label{genpol}
P_\mu(x)=\Pexp(-i\int^{l_\mu}_0 A_\mu(x)dx_\mu)\,\Omega_{e^{(\mu)}}(x)\quad.
\ee
In the presence of a space twist the $P_i$
are not arbitrary, but uniquely determined (up to some discrete transformations)
by the twist. This can be seen as follows. Since the configuration is 
exponentially localised in time, it goes to a pure gauge at $\pm\infty$. 
Now one can choose a gauge in which $A_i(t=\infty)=0$, fixing $\xi_i=0$
in Eq.~(\ref{abel}), for which the twist
matrices $\Omega_{e^{(i)}}$ are constant and the holonomy is given precisely by
these matrices. The fact that $\Omega_{e^{(1)}}$ and $\Omega_{e^{(3)}}$ anticommute (as a
result of the twist), implies they can be brought to the form $i\sigma_1$ 
and $i\sigma_3$, respectively. On the other hand, the constant twist matrix 
for the $2$ direction has to commute with these Pauli matrices and therefore 
must be $\pm I$. This fixes the Polyakov loops for the $Q=1$ space twisted 
instanton at both $t=\infty$ and $t=-\infty$ to be $\half\Tr P_1=\half
\Tr P_3=0$ and $\half\Tr P_2=\pm 1$, depending on whether we choose the 
twist matrix in direction 2 to be $\pm I$. The location of the caloron 
constituents are determined by the holonomies to be at $\vec{z}=(5,0,5)$, and 
the one displaced by $\half\vec{m}/l_s$ from it, i.e. $\vec{z}=(5,5,5)$.

We have generated two such configurations with time separations between the
$Q=\half$ instanton constituents given by $d_0=0.622$ and $d_0=0.786$.
In Figs.~7 and 8 we compare the Nahm transformed results with the infinite
volume analytic caloron solutions and the holonomy fixed to $\omega=d_0/2$.
Plotted is the action density along the line joining the two monopoles.
The comparison is very good taking into account that there are no free
parameters.  The mismatch at the peaks can be attributed to the fact that
the data points correspond to  a finite ratio $\l_t/l_s=1/10$,  which is zero for  the analytic result. 
A direct comparison with numerical data at this fixed ratio is difficult  due to
its small  value, and seems unnecessary given the agreement shown in Figs. 7-8.

\section{Conclusions}

By now our knowledge of self-dual configurations on the torus has come
from various sources. Apart from general existence proofs  and consequences
of the index theorem, information in this field has arisen from both numerical
studies based on lattice gauge theories and analytical results obtained
on manifolds having compact and non-compact directions. Most of the interest
and information refers to self-dual configurations on  spatially symmetric tori
of size $l_t \times l_s^3$ for large  and small aspect ratios, known as torus
instantons ($l_s/l_t \ll 1$) and torus calorons ($l_s/l_t \gg 1$). In this
paper we have studied the Nahm transformation that relates $SU(N)$ self-dual 
gauge fields of charge $Q$ with twisted boundary conditions on a torus, 
to $SU(N_0Q)$ self-dual gauge fields of charge $N/N_0$ on the dual torus.
Here $N_0$
is determined by the twist. We found that torus calorons and instantons
are dual to each other. This provides information on their respective moduli
spaces, which is fully consistent with the structure  suggested by numerical
studies.  In particular, it allows us to understand the findings of 
Ref.~\cite{Calo}. The most notable result is the  duality observed between 
holonomies and the location of constituent structures (BPS monopoles or twisted
instantons), as suggested by analysis of the Nahm transformation of
configurations living in $T^3\times R$ and $R^3\times S^1$. We showed how the 
abelian Nahm connections with singularities obtained in this case, result 
from the collapse of the non-abelian core of the constituents into point-like
singularities, which act as sources of the surviving abelian field.
Thus, all the information fits  nicely into a unified picture.
  
Our results might be of help in different respects.
Recently the ($T$) dualities discussed here have played important roles in
D-brane and string compactifications and our findings concerning the
relation between holonomies and constituent positions is an interesting one 
also in this context. For the low charges studied here we saw that the basic 
constituent is a twisted instanton, with charge $Q=\half$ in $SU(2)$, which 
shows up as a BPS monopole at finite temperature. We have shown 
that this fractional instanton maps onto itself  under the Nahm duality
transformation, a property guaranteed by the fact that for $SU(2)$ the twist 
is preserved under the duality transformation and the only freedom is the 
position of these twisted instantons. It would be interesting to see in which
sense these constituents do continue to play an important role at
higher charges. Finally, we hope that the work presented here will
lead to a more complete analytic understanding.

\section*{Appendix}

In this appendix we will provide the proofs of the basic ingredients of the
flavour construction of Nahm's  transformation for non-trivial twisted  
boundary conditions ($\vec{k}$, $\vec{m}\ne\vec{0}) $. In particular, we 
will present in a basis independent way the derivation of the characterisation 
for the Nahm-dual torus and its corresponding dual twist. This complements 
the results of Ref.~\cite{Ntbc}. 

Our starting point is the set of Weyl zero-modes $\Psi_z^{i\alpha}(x)$ 
satisfying the boundary conditions of Eq.~(\ref{boundary2}). Consider now 
the behaviour of these zero-modes under translation of the variables $z$. 
It is easy to see that given a solution $\Psi_{z}(x)$ of the Weyl equation 
one can construct a new solution of the Weyl equation as follows:
\be
\chi^{(y)}_z(x)\equiv e^{-2\pi i y(x)}\,\Psi_{z+y}(x)\quad,
\ee
for any element $y\in \tilde{R}^4$
(as usual the space of linear forms $\tilde R^4$ can be identified 
with $R^4$, and $y(x)\equiv y \cdot x$). 
Applying 
this to an orthonormal set  of solutions of the Weyl equation one obtains 
a new set. This set satisfies different boundary conditions, obtained from 
Eq.~(\ref{boundary2}) by replacing the twist eating matrices $\Gamma(a)$,
\be
\Gamma(a) \longrightarrow \Gamma'(a)= e^{2 \pi i y(a)}\,\Gamma(a)\quad.
\ee
We thus see that there is a correspondence between translations in $z$ and 
different choices of twist-eating solutions. 

In general the matrices $\Gamma'(a)$ constitute an inequivalent set of
solutions to the original $\Gamma(a)$. However, for special values of $y$ the 
two sets are unitarily equivalent (related by a similarity transformation). 
For those values of $y$, denoted by $\hat a$, there exist $U(N_0)$ matrices 
$\tilde\Gamma(\hat a)$ satisfying
\be
\label{dual}
\tilde\Gamma(\hat a)\,\Gamma(a)=e^{2\pi i\hat
a(a)}\,\Gamma(a)\,\tilde\Gamma(\hat a)\quad.
\ee
This formula shows a nice duality between both sets of $U(N_0)$ matrices. 
Before investigating for which $\hat a$ these equations have solutions, we 
want to discuss its consequences. For all values $\hat a$, for which 
Eq.~(\ref{dual}) has solutions, we can construct the functions
\be
e^{-2\pi i\hat a(a)}\,\Psi_{z+\hat a}^{j\alpha}(x)\,\tilde\Gamma^{ji}(\hat a)
\ee
which form a new orthonormal set of solutions of the Weyl equation with the
{\em same} boundary conditions Eq.~(\ref{boundary2}). Thus, they can be 
expressed as a unitary combination of the original set of zero-modes. We 
arrive at the main formula:
\be
\label{main1}
\Psi_{z+\hat a}^{i\alpha}(x)= e^{2\pi i\hat a(a)}\,\tilde\Gamma^*_{ij}(\hat a)\,
\Psi_{z}^{j\beta}(x)(\hat{\Omega}_{\hat a}^{\dagger}(z))^{\beta\alpha}\quad.
\ee
From this one can deduce that for all $\hat a$ the Nahm transformed gauge 
field satisfies
\be
\label{mainform}
\hat A_{\mu}(z+\hat a)=[\hat\Omega_{\hat a}(z)]\,\hat A_\mu(z)
\ee
One easily sees that the set of $\hat a$ define a lattice $\hat\Lambda$, which 
we will call the Nahm-dual lattice. It is a sublattice of $\tilde\Lambda$,
the lattice dual to $\Lambda$. We define the Nahm transformation of the 
original self-dual gauge field with twist as the restriction of $\hat A$
to the Nahm-dual torus $R^4/\hat\Lambda$. This is a $U(N_0Q)$ self-dual gauge 
field with topological charge $\hat Q=N_0N/|\tilde\Lambda/\hat\Lambda|$,
where $|\tilde\Lambda/\hat\Lambda|$ stands for the number of unit cells
of $\hat\Lambda$ that fit in a unit cell of $\tilde\Lambda$, or the number
of elements in $\hat\Lambda$(mod $\tilde\Lambda$). The matrices 
$\hat\Omega_{\hat a}(z)$ are the corresponding twist matrices.

Now let us investigate the solutions to Eq.~(\ref{dual}) and the
characterisation of $\hat\Lambda$. For that it is essential to use some 
properties of the twist eating solutions. One can prove~\cite{RevT} 
that the set of matrices $\Gamma(a)$ contains a set of $N_0^2$ linearly 
independent matrices, which form a basis of the space of all $N_0\times N_0$
matrices. Actually, to construct this basis one has to select a representative 
of the quotient lattice $\Lambda/\Lambda_0$. Its order is precisely $N_0^2$. 
The sublattice $\Lambda_0$ can also be defined as the one associated to the
matrices $\Gamma(a)$ commuting with all other $\Gamma(a')$ for any 
$a'\in\Lambda$. We call it the center lattice. Using this fact one can, after 
some effort, prove that indeed the matrices $\tilde\Gamma(\hat a)$ do coincide, 
up to a  phase factor, with the original matrices $\Gamma(a)$. 
In more mathematical terms, we can say that there exists a mapping $\phi$ from
$\hat\Lambda$ to $\Lambda$ such that
\be
\label{rel}
\tilde\Gamma(\hat a)=e^{ i\varphi(\hat a)}\,\Gamma(\phi(\hat a))\quad.
\ee
where $\varphi(\hat a)$ is an arbitrary phase.  
From this we can identify $\hat\Lambda$. Let $a$ be an element of $\Lambda_0$, 
then it commutes with all matrices $\Gamma(a')$ for any $a\in\Lambda'$, this 
includes the matrices $\tilde\Gamma(\hat a)$. Inspecting Eq.~(\ref{dual})
we conclude that $\hat{a}(a)\in Z$ for any $a\in\Lambda_0$. Therefore
$\hat\Lambda=\tilde\Lambda_0$, the Nahm dual lattice coincides with the dual 
lattice of the center lattice. One can also see that the center lattice of 
$\hat\Lambda$ coincides with $\tilde\Lambda$. From here it is immediate to 
see that the Nahm-dual of the Nahm-dual gives back the original lattice. 

There is one extra ingredient which has to be worked out, namely what is the 
Nahm dual twist. This follows from Eq.~(\ref{rel}) and Eq.~(\ref{twisteaters}).
We have
\be
\tilde\Gamma(\hat a)\,\tilde\Gamma(\hat a')=e^{-2\pi i\hat\NN(\hat a,
\hat a'))}\,\tilde\Gamma(\hat a')\,\tilde\Gamma(\hat a)
\ee
with
\be
\label{twistF}
\hat\NN(\hat a,\hat a')=-\NN(\phi(\hat a),\phi(\hat a'))\quad .
\ee
Eq.~(\ref{main1}) is only consistent provided the Nahm-dual twist matrices 
$\hat\Omega_{\hat a}(z)$ satisfy twisted boundary conditions analogous to 
Eq.~(\ref{tbc1}) with the skew symmetric form $\NN$ replaced by 
$\hat\NN$, the dual twist form.

\section*{Acknowledgements}
We are grateful to Tam\'as Kov\'acs and \'Alvaro Montero for useful discussions.
This work was supported in part by a grant from ``Stichting Nationale Computer 
Faciliteiten (NCF)'' for use of the Cray Y-MP C90 at SARA. A. Gonz\'alez-Arroyo
and C. Pena acknowledge financial support by CICYT  under grant AEN97-1678. We 
acknowledge Centro de Computaci\'on Cient\'{\i}fica at UAM for the availability of 
computer resources. M. Garc\'{\i}a P\'erez acknowledges financial support by 
CICYT and warm hospitality at the Instituut Lorentz while part of this work 
was developed.

\end{document}